\documentclass[twocolumn,prl,superscriptaddress,amsmath,amssymb]{revtex4}
\usepackage{graphicx}
\usepackage{bm}
\usepackage{hyperref}
\hypersetup{colorlinks=true,allcolors=blue}

\begin{document}

\title{Exotic Acoustic-Edge and Thermal Scaling in Disordered Hyperuniform Networks }
%with Compensated Kernels
% not sure if it is really universal, need to polish a little more

\author{Yang Jiao}
\affiliation {Materials Science and Engineering, Arizona State University, Tempe, AZ 85287} \affiliation{Department of Physics, Arizona State University, Tempe, AZ 85287}

\date{\today}

\begin{abstract}
We develop a first-principles theory for the vibrational density of states (VDOS) and thermal properties of network materials built on stationary correlated disordered point configurations. For scalar (mass--spring) models whose dynamical matrix is a distance-weighted graph Laplacian, we prove that the limiting spectral measure is the pushforward of Lebesgue measure by a Fourier symbol that depends only on the edge kernel \(f\) and the two-point statistics \(g_2\) (equivalently the structure factor \(S\)). For hyperuniform systems with small-$k$ scaling \(S(k)\sim k^\alpha\) and compensated kernels, {the VDOS exhibits an algebraic \emph{pseudogap} at low frequency, \(g(\omega)\sim \omega^{\,2d/\beta-1}\) with \(\beta=\min\{4,\alpha+2\}\), which implies a low-temperature specific heat \(C(T)\sim T^{\,2d/\beta}\) and a heat-kernel decay \(Z(t)\sim t^{-d/\beta}\), defining a spectral dimension \(d_s=2d/\beta\).} This hyperuniformity-induced algebraic edge depletion could enable novel wave manipulation and low-temperature applications. Generalization to vector mechanical models and implications on material design are also discussed.
\end{abstract}

%at the edge is in contrast to vector mechanical models (full dynamical matrices) that can display true low-frequency gaps in specialized regimes

%In the stealthy hyperuniform (SHU) limit where \(S(k)=0\) for \(|k|<K\) (effectively \(\alpha\to\infty\)), the non-analytic contribution vanishes and the edge exponent reverts to Debye (\(\beta=2\)), i.e., \(g(\omega)\propto \omega^{d-1}\) and \(C(T)\propto T^{d}\), with a crossover near \(\omega_K\sim \sqrt{C_2}\,K\).

%We position these results relative to numerical studies of ``hyperuniform glasses'' and outline parameter-free tests based on measured \(S(k)\).

\maketitle

%Disordered solids with \emph{hyperuniform} microstructure---suppressed long-wavelength density fluctuations, \(S(k)\to 0\) as \(k\to 0\)---exhibit unusual transport and wave phenomena \cite{TorquatoStillinger03,Torquato18}. 

Disorder hyperuniform (DHU) many-body systems lack conventional long-range order as in an amorphous material, yet they possess a ``hidden order'' manifested as complete suppression of normalized infinite-wavelength density fluctuations like crystals \cite{To03, To18a}. Recently, a wide spectrum of equilibrium \cite{To15, Ba09} and non-equilibrium \cite{Ga02, Do05, Za11a} many-body systems, in both classical \cite{Ku11, Hu12, Dr15, chen2021multihyperuniform, zhang2023approach, Ch18b} and quantum mechanical \cite{Fe56, Ge19quantum, sakai2022quantum, Ru19, Sa19, Zh20, Ch21, chen2025anomalous} varieties, have been identified to possess the property of disordered hyperuniformity. Other examples include certain biological systems \cite{Ji14, ge2023hidden, liu2024universal}, driven non-equilibrium systems \cite{He15, Ja15, We15, salvalaglio2020hyperuniform, nizam2021dynamic, zheng2023universal, wang2025hyperuniformprl}, active-particle fluids \cite{Le19, lei2019hydrodynamics, huang2021circular, zhang2022hyperuniform, oppenheimer2022hyperuniformity}, and dynamic random organizing systems \cite{hexner2017noise, hexner2017enhanced, weijs2017mixing, wilken2022random}. Novel DHU materials have been engineered that can possess superior properties compared to their crystalline counterpart, such as high-degree of isotropy and robustness against defects \cite{Fl09, klatt2022wave, Zh16, Ch18a, torquato2021diffusion, Xu17, Le16, yu2023evolving}.

Hyperuniformity is characterized by a local number variance $\sigma_N^2(R)$ associated with a spherical window of radius $R$ in $\mathbb{R}^d$ that grows more slowly than the window volume in the large-$R$ limit \cite{To03, To18a}, i.e., $\lim_{r\rightarrow \infty} \sigma_N^2(R)/R^d = 0$. Equivalently, the static structure factor vainshes in the infinite-wavelength (or zero-wavenumber) limit, i.e., $lim_{|{\bf k}|\rightarrow 0}S({\bf k}) = 0$, where ${\bf k}$ is the wavenumber and $S({\bf k})$ is related to the pair-correlation function $g_2({\bf r})$ via $S({\bf k}) = 1+\rho \int e^{-i{\bf k}\cdot {\bf r}}[g_2({\bf r}) - 1]d{\bf r}$ and $\rho = N/V$ is the number density of the system. For statistically isotropic systems, the structure factor only depends on the wavenumber $k = |{\bf k}|$. The small-$k$ scaling behavior of $S(k)$, i.e., $S({k}) \sim k^\alpha$ determines the large-$R$ asymptotic behavior of $\sigma_N^2(R)$, based on which all DHU
systems can be categorized into three classes:
$\sigma_N^2(R) \sim R^{d-1}$ for $\alpha>1$ (class I); $\sigma_N^2(R)
\sim R^{d-1}\ln(R)$ for $\alpha=1$ (class II); and $\sigma_N^2(R)
\sim R^{d-\alpha}$ for $0<\alpha<1$ (class III), where $\alpha$ is the hyperuniformity exponent  \cite{To18a}.

%Designer DHU materials have also been successfully fabricated or
%synthesized using different techniques \cite{ref36, ref37}.

Recent numerical studies on hyperuniform glasses and designed disordered materials \cite{mizuno2017continuum, hu2022origin, xu2024low, wang2025hyperuniform} report depleted low-frequency modes, modified heat capacities, and in some models apparent low-frequency gaps \cite{zhuang2024vibrational}. In contrast to ordered network systems \cite{lubensky2015phonons, mao2018maxwell}, what has been missing for disordered systems is a compact analytic relation that predicts the low-frequency vibrational density of states (VDOS) and associated thermal exponents directly from the structural statistics of the disordered underlying point configuration (e.g., distribution of the particle centers).

%Here, we develop a first-principles theory for the vibrational density of states (VDOS) and thermal properties of network materials built on stationary correlated disordered point configurations.

Here we provide such a relation for scalar mass--spring networks by developing a first-principles theory. Specifically, the VDOS is a pushforward of Lebesgue measure by a symbol \(\Lambda_L(k)\) of the scalar dynamical Laplacian $L$, determined only by the edge kernel \(f\) and pair statistics \(g_2\) (or equivalently structure factor $S$). For compensated kernels (see definition below), hyperuniformity yields a universal algebraic \emph{pseudogap} at the acoustic edge with exponent fixed by the small-\(k\) scaling of \(S(k)\). Our framework bridges random geometric graph theory and the statistical mechanics of hyperuniform point processes, and offers parameter-free predictions testable against numerical results and inspires novel material design.

\emph{\bf Scalar dynamical matrix as a distance-weighted Laplacian.}
Let \({\bf x}_1,\dots,{\bf x}_N\in\Omega\subset\mathbb{R}^d\) be points from a stationary, isotropic process of number density \(\rho=N/Vol(\Omega)\) and pair correlation function $\rho^{(2)}({\bf x}, {\bf y}) = \rho^2 g_2(|{\bf x}-{\bf y}|) = \rho^2 g_2(r)$, where $r = |{\bf x}-{\bf y}|$ is the Euclidean distance between point ${\bf x}$ and ${\bf y}$. Consider a nonnegative, rapidly decaying radial kernel \(f(r)\) defining weighted edges between sites, which also defines a graph based on the point configuration. Examples of $f(r)$ include inverse power functions, Gaussian, or exponential kernels. The connectivity/adjacency matrix $A$ has entries 
\begin{equation}
A_{ij}=
\begin{cases}
f(|{\bf x}_i-{\bf x}_j|), & i\neq j,\\[2pt]
0, & i=j.
\end{cases}
\end{equation}
In this work, we focus on the ``scalar elasticity'' problem based on the graph (i.e., a network of scalar ``springs''), i.e., 
\begin{equation}
    L u_n = \lambda u_n
\end{equation}
where $L$ is the (mass-normalized) scalar dynamical matrix and $\lambda_n = \omega_n^2$ is the eigen value of the matrix, with $\omega_n$ being the vibration frequencies. We note in this setting, $L$ is also the graph Laplacian, i.e., $L = D - A$, with $D = Diag(d_1, \ldots, d_n)$ and $d_i = \sum_j f(|{\bf x}_i-{\bf x}_j|)$ \cite{Chung1997SGT, DoyleSnell1984}. Thus the entries of the Laplacian is given by
\begin{equation}
L_{ij}=
\begin{cases}
-\;f(|{\bf x}_i-{\bf x}_j|), & i\neq j,\\[2pt]
\sum_{m\neq i} f(|{\bf x}_i-{\bf x}_m|), & i=j.
\end{cases}
\end{equation}
$L$ is symmetric, positive semidefinite and its eigenvalues \(\lambda_n\ge 0\) are related to mode frequencies by \(\omega_n^2=\lambda_n\). Denote by \(\varphi_L(\lambda)\) the limiting eigenvalue density (per site in the infinite $N$ limit) of \(L\), the vibrational density of states (VDOS) is given by \cite{AshcroftMermin1976, LifshitzGredeskulPastur1988}
\begin{equation}
g(\omega)=2\omega\,\varphi_L(\lambda)\big|_{\lambda=\omega^2}.
\label{eq:VDOSmap}
\end{equation}

%With \(\rho^{(2)}(x,y)=\rho^2 g_2(|x-y|)\) and ,

\begin{figure*}[ht]
\begin{center}
$\begin{array}{c}\\
\includegraphics[width=0.85\textwidth]{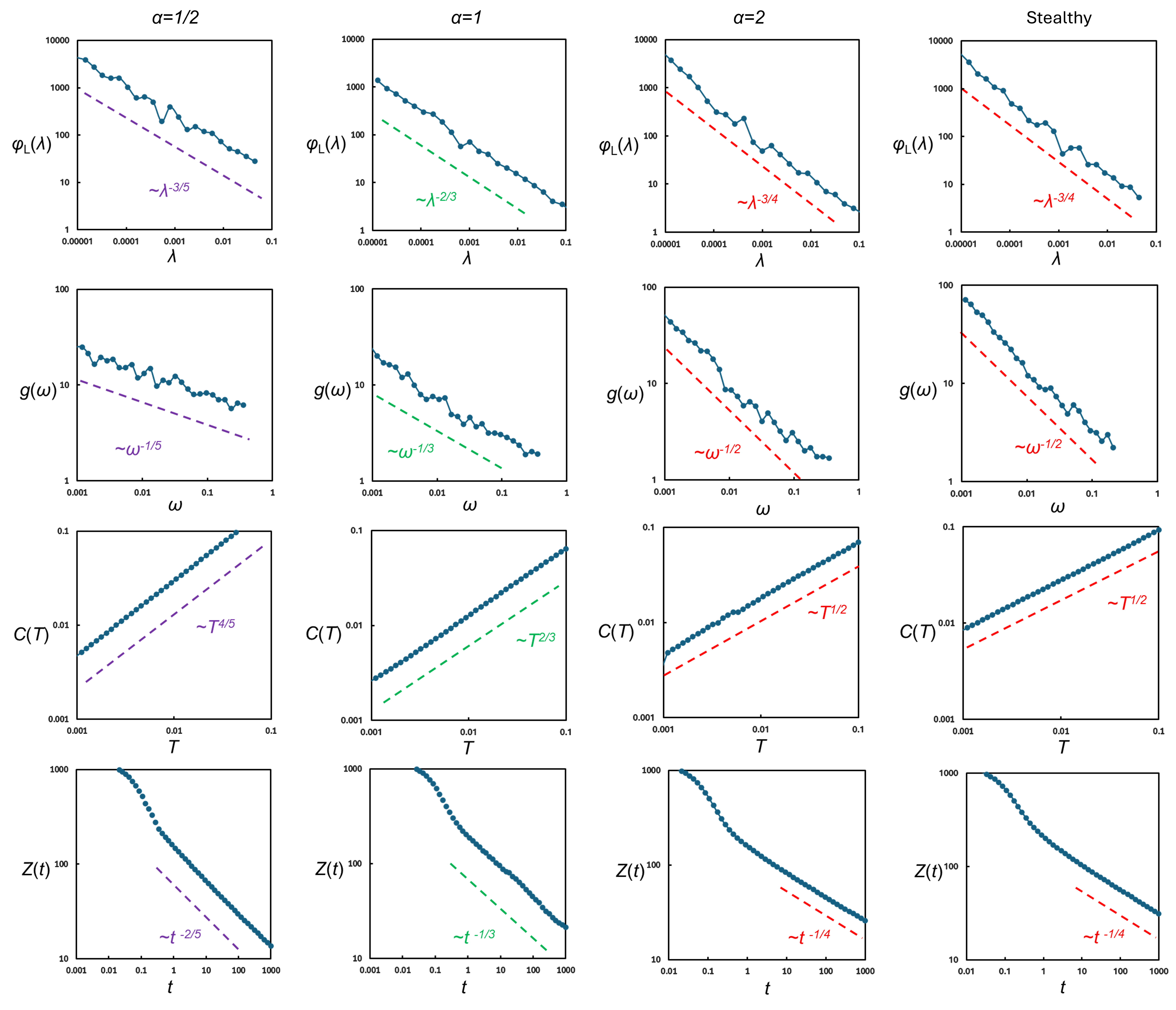}
\end{array}$
\end{center}
\caption{Comparisons of theoretical predictions for eigen value distribution $\varphi_L(\lambda)\sim \lambda^{d/\beta-1}$, VDOS $g(\omega) \sim \omega^{2d/\beta-1}$, heat capacity $C(T)\sim T^{2d/\beta}$ (per node, with unit $k_B$) and heat kernel $Z(t) \sim t^{-d/\beta}$ with numerical results of a variety of hyperuniform systems with $\alpha = 1/2$, 1, 2 and stealthy systems in 1D with $N=10,000$ points in a unitary periodic box and the compensated kernel (\ref{eq_ck}) with $a = 0.002$ and $b=0.004$. The scaling behaviors of the reported quantities are not sensitive to choice of $a$ and $b$ values. The results are obtained by averaging over 10 independent realizations for each case.} \label{fig1}
\end{figure*}

\emph{\bf DOS from two-point statistics.} Consider the static structure factor of the point configuration \(S(k)=1+\rho\,\widehat{h}(k)\) where $\widehat{h}(k)$ is the Fourier transform of the total correlation function $h(r) = g_2(r) - 1$, the operator \(L\) converges (in the thermodynamic limit, in the sense of empirical spectral measures) to a translation-invariant convolution operator with symbol (i.e., the Fourier transform of the operator) \cite{BelkinNiyogi2007, GarciaTrillosSlepcev2018, adhikari2022spectrum}
\begin{equation}
    \Lambda_L(k) =\bar d-\Lambda_A(k)
\end{equation}
where 
\begin{equation}
    \bar d=\Lambda_A(0)=\rho\!\int f(r)g_2(r)\,d^dr
\end{equation}
and 
\begin{equation}
   \Lambda_A(k)=\rho\,\widehat{\,f\,g_2\,}(k)
=\rho\,\hat f(k)+\frac{1}{(2\pi)^d}\big(\hat f*(S-1)\big)(k), 
\end{equation}
and $\Lambda_A(k)$ is the symbol of the adjacency matrix operator. For any bounded continuous test function \(\phi\), the convergence of the empirical spectral measures is stated as \cite{ReedSimonIV1978, BottcherSilbermann1999}:
\begin{equation}
\lim_{N\to\infty}\frac{1}{N}\,\mathbb{E}\big[\mathrm{Tr}\,\phi(L)\big]
=\frac{1}{(2\pi)^d}\int \phi\big(\Lambda_L(k)\big)\,d^dk,
\label{eq:pushforward}
\end{equation}
and the limiting density of eigenvalues is
\begin{equation}
\varphi_L(\lambda)=\frac{1}{(2\pi)^d}\int \delta\!\big(\lambda-\Lambda_L(k)\big)\,d^dk.
\label{eq:DOS}
\end{equation}
where $\delta(\cdot)$ is the Dirac delta function. Equations \eqref{eq:VDOSmap}--\eqref{eq:DOS} connects vibrational spectra prediction to two-point information.

%\textcolor{red}{should not talk about Poisson, directly introduce the DHU scaling, and provide the expansion, then introduction compensated kernel}

%make a note that for crystal, it is different mechansim leading to the same behavior. 

\emph{\bf Hyperuniform acoustic-edge scaling.}
We now consider a power-law small-$k$ scaling of the structure factor, i.e., \(S(k)\sim A_H |k|^\alpha\) as \(k\to 0\) with \(\alpha>0\), and assume that \(\hat f\) is analytic near \(k=0\). Then the non-analytic contribution from \(S-1\) affect the symbol near the acoustic edge as follows (see SI for details):
\begin{equation}
\Lambda_L(k)=C_2 |k|^2 + C_{2+\alpha}\,|k|^{2+\alpha} + C_4 |k|^4+o(|k|^4),
\label{eq:symbolEdge}
\end{equation}
where $C_2 = \frac{\rho}{2d}M_2$ and $C_4 = \frac{\rho}{8d(d+2)}M_4$ and $M_n = \int_{\mathbb{R}^d}r^n f(r)d^dr$; $C_{2+\alpha}$ is the coefficient associated with the $S(k)$ contribution, which is given in SI. 

It can be seen from Eq.(\ref{eq:symbolEdge}) that for typical kernels (Gaussian, exponential, inverse power-law), the $k^2$ term is dominating. Keeping only the leading term $\Lambda_L(k) \approx \frac{\rho M_2}{2d}k^2$ near the edge, Eq. (\ref{eq:DOS}) for small $\lambda$ yields
\begin{equation}
    \varphi^{D}_L(\lambda)\approx\frac{S_{d-1}}{2(2\pi)^d}(\frac{\rho}{2d}M_2)^{-d/2}\lambda^{d/2-1}
\end{equation}
where $S_{d-1}=2\pi^{d/2}/\Gamma(d/2)$ is the $d$-dimensional solid angle. For VDOS, this yield
\begin{equation}
    g^{D}(\omega)=2\omega\,\varphi_L(\omega^2) \sim \omega^{d-1}
\end{equation}
which is the universal Debye exponent for the low-frequency state \cite{AshcroftMermin1976, kittel2018introduction}. In this case, the effect of hyperuniformity ($\alpha >0$) is masked by the analyticity of the kernel function. 

%We also note that crystalline solids possess the same Debye exponent at low frequency because their acoustic phonons possess linear dispersion relation \cite{AshcroftMermin1976, kittel2018introduction}. 

Here we consider the so-called {\it compensated kernels}, i.e., one that satisfies 
\begin{equation}
    M_2 = \int_{\mathbb{R}^d}r^2 f(r)d^dr = 0
\end{equation}
and $M_4>0$, which contributes to the Laplacian symbol with $k^4$ term. Examples of such systems include bending-dominated elastic medium, i.e. the discrete analog of the Kirchhoff–Love plate or Euler–Bernoulli beam \cite{timoshenko1959theory} and 2D materials \cite{nika2017phonons}. An example of such compensated kernel is
\begin{equation}
f_{c}(r)=e^{-(r/a)^2}-\eta\,e^{-(r/b)^2},
\label{eq_ck}
\end{equation}
where $\eta=\Big(\frac{a}{b}\Big)^{d+2}$ and $0<a<b$ are length scale parameters. 

For a Poisson distribution of points, $S(k) = 1$ and the spectra symbol of the adjacency and Laplacian operators reduce to $ \Lambda_A(k)=\rho \widehat{f}(k)$ and $\Lambda_L(k)= \bar{d} -\Lambda_A(k) =  \rho [\widehat{f}(0) - \widehat{f}(k)]$, respectively. With the compensated kernel Eq.(\ref{eq_ck}), the leading term is $\Lambda_L(k) \approx C_4 k^4$, which yields
\begin{equation}
\varphi^P_L(\lambda)
\;\sim\; \frac{S_{d-1}}{4(2\pi)^d} C_4^{-\frac{d}{4}}\,\lambda^{\,\frac{d}{4}-1}
\label{eq:lamScalingPoisson}
\end{equation}
and 
\begin{equation}
g^P(\omega)\;\sim\; \frac{S_{d-1}}{2(2\pi)^d}C_4^{-d/4}\;
\omega^{\,\frac{d}{2}-1} \sim \omega^{\,\frac{d}{2}-1}
\label{eq:VDOSScaling}
\end{equation}

For hyperuniform systems ($0<\alpha<2$) with compensated kernels, the leading order term is $\Lambda_L(k) \approx C_{2+\alpha} k^{2+\alpha} + o(k^4)$. Plugging this into Eq. (\ref{eq:DOS}) yields the following scaling of the edge states:
\begin{equation}
\varphi^H_L(\lambda)
\;\sim\; \frac{S_{d-1}}{(2\pi)^d}\frac{1}{\beta}\,C_\beta^{-d/\beta}\,\lambda^{\,\frac{d}{\beta}-1}
\label{eq:lamScaling}
\end{equation}
and, via \eqref{eq:VDOSmap}, the VDOS near the edge is given by
\begin{equation}
g^H(\omega)\;\sim\; \frac{S_{d-1}}{(2\pi)^d}\frac{2}{\beta}\,C_\beta^{-d/\beta}\;
\omega^{\,\frac{2d}{\beta}-1} \sim \omega^{\,\frac{2d}{\beta}-1}
\label{eq:VDOSScaling}
\end{equation}
where $\beta=\min\{4,2+\alpha\}$. 

Eq. (\ref{eq:VDOSScaling}) indicates that, for a hyperuniform system in ($d\ge 2$) with hyperuniformity exponent $0<\alpha<2$, the spectrum touches zero, yet the density of states near the edge vanishes algebraically with a stronger power law than the Poisson and general Debye systems. For $d=1$, a stronger depletion at the edge is also manifested, e.g., as the slower diverging the zero-$k$ states. Because \(\Lambda_L(k)\) is a continuous scalar function on a connected Brillouin zone, the spectrum of \(L\) is the essential range of \(\Lambda_L\), i.e., a \emph{single connected band} touching 0 at \(k=0\). Thus the scalar model produces an \emph{algebraic pseudogap} (depletion) at the edge, not a true forbidden interval.

%\textcolor{red}{Figure \ref{fig1} shows the comparison of our theory to numerical data for a variety of hyperuniform systems with varying $\alpha$.}

%We numerically verify the theoretical predictions using 1D systems including Poisson and a variety of disordered hyperuniform distributions of points ($\alpha = 0.5$, 1, 1.5, 2 and stealthy) with $N = 10,000$. For each system, 10 configurations are used for ensemble average. The DHU configurations are generated using the collective coordinate approach \cite{Ba08}. Figure \ref{fig1}(a) and (b) respectively compares our theory to numerical results for the edge behavior of $\varphi_L(\lambda)$ and $g(\omega)$ and excellent agreement is observed for all systems.

We numerically test our theoretical predictions using one-dimensional point configurations, including Poisson (see Supporting Information) and several classes of disordered hyperuniform (DHU) systems with exponents $\alpha = 1/2,1,2$, as well as stealthy patterns, each containing $N=10{,}000$ points. For every case, ensemble averages are obtained over 10 independent realizations. The DHU configurations are generated via the collective coordinate method \cite{Ba08}. Figures~\ref{fig1} top two rows respectively compare the theoretical predictions with numerical results for the edge behavior of $\varphi_L(\lambda)$ and $g(\omega)$, showing excellent agreement across all systems.

We note that the stronger power-law depletion of low-frequency modes in hyperuniform networks effectively trims the soft tail of the spectrum, which can significantly impact the elastic moduli, wave responses and vibrational noise floors. For elastic moduli, non-affine correction terms scale like $\int g(\omega)\Gamma(\omega)/\omega^2\,d\omega$; thus, depletion of low-$\omega$ modes lowers these corrections, pushing moduli closer to their affine (stiffer) values and reducing sample-to-sample variability. For wave response, the scarcity of ultra-soft modes, together with suppressed long-wavelength density fluctuations, reduces large-scale scattering/attenuation in the acoustic range. Moreover, quantities weighted by $1/\omega^2$ (e.g.\ thermal displacement noise) decrease as the low-$\omega$ DOS is thinned, enabling lower noise and potentially higher precision engineering.

\emph{\bf Hyperuniform thermal exponents.} Consider the internal energy of harmonic oscillators, i.e.,
\begin{equation}
\begin{array}{c}
     U(T)=\int_0^\infty \hbar\omega\,\frac{1}{e^{\hbar\omega/k_BT}-1}\,g(\omega)\,d\omega\;\\
     \sim\; A_\beta \int_0^\infty \hbar\omega\,\frac{\omega^{p}}{e^{\hbar\omega/k_BT}-1}\,d\omega
\end{array}
\label{eq_UT}
\end{equation}
where $p=\frac{2d}{\beta}-1$. Simplifying Eq. (\ref{eq_UT}) yields the low-$T$ scaling (see SI for details):
\begin{equation}
    U(T) \sim T^{2d/\beta + 1}.
\end{equation}
The heat capacity $C(T)$ is then obtained by differentiating the internal energy, i.e.,
\begin{equation}
    C(T)=\frac{dU}{dT}\sim T^{2d/\beta},\qquad
\beta=\min\{4,2+\alpha\}
\end{equation}
For Poisson systems, we have $\beta=4$, which leads to the low-$T$ scaling of heat capacity $C^P(T)\sim T^{d/2}$. Hyperuniform systems with $0<\alpha<2$ have $\beta=2+\alpha<4$, giving
\begin{equation}
    C^H(T)\ \sim\ T^{\,2d/(2+\alpha)},
\end{equation}
with a \emph{higher} power than $T^{d/2}$ and therefore a much stronger low–$T$ suppression. 
In $d=3$, e.g.\ $\alpha=1$ (jammed particle packings \cite{Do05}) yields $C^H(T)\!\sim\!T^{2}$ (vs.\ $T^{3/2}$ in Poisson systems), implying orders-of-magnitude reduction at millikelvin–kelvin scales. Therefore, such hyperuniform materials can enable fast thermal response (for fast thermometry and pulsed operation), thermal management by design (e.g., engineering $S(k)$ to target specific $C(T)$ curves for cryogenic subsystems), and enhance sensitivity and bandwidth at low $T$ calorimetry and bolometry. Figure~\ref{fig1} third row compares the theoretical predictions with numerical results for the low-$T$ scaling of the heat capability $C(T)$ and excellent agreement across all systems is observed.

%\textcolor{red}{Need to define the heat transfer problem on network}

Finally, we note that the Laplacian $L$ also governs the diffusion/heat transport on network, i.e., 
\begin{equation}
d{\bf T}(t)/dt = -L {\bf T}(t) \quad {\bf T}(t) = e^{-Lt}{\bf T}(0)
\end{equation}
Thus, the same spectra used for vibrations also controls relaxation on networks. We consider the heat trace 
\begin{equation}
    Z(t)=\mathrm{Tr}\,e^{-tL},
\end{equation}
which governs the relaxation in the system. For large $t$, the heat trace possesses the following form
\begin{equation}
    Z(t)=\int_0^\infty e^{-t\lambda}\varphi_L(\lambda)\,d\lambda
\end{equation}
which is dominated by small $\lambda$. Plugging in Eq. (\ref{eq:lamScaling})
\begin{equation}
\begin{array}{c}
        Z(t)\ \sim\ \frac{S_{d-1}}{(2\pi)^d}\frac{1}{\beta}\,C_\beta^{-d/\beta}
\int_0^\infty e^{-t\lambda}\,\lambda^{\,\frac{d}{\beta}-1}\,d\lambda \\
= \frac{S_{d-1}}{(2\pi)^d}\frac{1}{\beta}\,C_\beta^{-d/\beta}\,
\Gamma\!\Big(\frac{d}{\beta}\Big)\,t^{-d/\beta}.
\end{array}
\end{equation}
which indicates the large-$t$ scaling \(Z(t)\sim c\,t^{-d/\beta}\), defining a spectral dimension \(d_s=2d/\beta\). Figure \ref{fig1} fourth row shows the comparison of our theory to numerical data, which again shows excellent agreement.

%In the non-hyperuniform (Poisson-like) case, \(\beta=2\) and \eqref{eq:VDOSScaling} reduces to the Debye law \(g(\omega)\propto \omega^{d-1}\) with \(C(T)\propto T^d\). 

For hyperuniform systems with \(0<\alpha<2\), \(\beta=2+\alpha<4\), $Z^H(t) \sim t^{-d/{(2+\alpha)}}$ decays faster than the Poisson systems with \(Z^P(t)\sim c\,t^{-d/4}\), implying quicker equilibration, shorter mixing times, and less thermal memory. This result is consistent with a recent study on the diffusion spreadability, characterizing transport in two-phase media, which indicates hyperuniform systems can achieve exponentially fast homogenization \cite{torquato2021diffusion}.

%\noindent\emph{Stealthy hyperuniformity} ($S(k)=0$ for $|k|<K$) restores the Debye \emph{exponent} ($\beta=2$) but still permits \emph{prefactor} control via $C_2=\tfrac{\rho}{2d}\!\int r^2 f(r)g_2(r)\,d^dr$ and a tunable crossover at $k\!\sim\!K$; non-stealthy hyperuniformity ($0<\alpha<2$) yields the genuinely stronger depletion ($\beta=\alpha<2$) that most benefits the effects above.

\emph{\bf Discussion.} For \emph{Stealthy hyperuniform} (SHU) system possessing \(S(k)=0\) for \(|k|<K\) (effectively \(\alpha\to\infty\)), the non-analytic contribution from \(S-1\) is absent and the analytic kernel term controls the edge. Hence \(\beta=4\) for compensated kernels, \(g(\omega)\propto \omega^{d/2-1}\) with \(C(T)\propto T^{d/2}\). 

%\textcolor{red}{Also need to comment on antihyperuniform systems.}

For {\it antihyperuniform} (AHU) point sets with $S(k)\sim k^{\alpha}$ and $-d<\alpha<0$, the correlation cusp dominates the analytic $k^2$ term [c.f. Eq.(\ref{eq:symbolEdge})] and renormalizes the small-$k$ symbol to a \emph{fractional order} $\Lambda_L(k)\sim k^{\gamma}$ with $\gamma=2+\alpha\in(0,2)$, even for generic short-range kernels (no compensation needed). By phase–space counting this yields non-Debye edge exponents:
$\varphi(\lambda)\sim \lambda^{d/\gamma-1}$,
$g(\omega)\sim \omega^{2d/\gamma-1}$,
$C(T)\sim T^{2d/\gamma}$,
and $Z(t)\sim t^{-d/\gamma}$.
Because $\gamma<2$, AHU media exhibit a \emph{stronger depletion} of acoustic-edge states than Debye, which is an apparently counter-intuitive result that follows from the more nonlinear $k\!\mapsto\!\omega$ mapping of a softer (fractional) operator, not from increased “order.” %(If $\alpha\le -d$, infrared divergences require a physical cutoff and modify the asymptotics.)

%Stealthiness still matters through the \emph{prefactor} \(C_2=\tfrac{\rho}{2d}\!\int r^2 f(r)g_2(r)\,d^dr\) and an intrinsic crossover scale \(\omega_K\sim \sqrt{C_2}\,K\), below which the Debye law holds exactly and above which features appear as \(S(k)\) turns on for \(|k|\gtrsim K\) (see SI for details). 

%For long-range kernels \(f(r)\sim r^{-(d+\sigma)}\) with \(0<\sigma<2\), the kernel non-analyticity dominates and \(\beta=\sigma\) even in the SHU limit.

%\emph{Band connectivity and ``no true gap'' in the scalar model.}---
%Because \(\Lambda_L(k)\) is a continuous scalar function on a connected Brillouin zone, the spectrum of \(L\) is the essential range of \(\Lambda_L\), i.e., a \emph{single connected band} touching 0 at \(k=0\). Thus the scalar model produces an \emph{algebraic pseudogap} (depletion), not a true forbidden interval. True low-frequency \emph{gaps} that appear in certain mechanical models arise from matrix-valued dynamical matrices (vector elasticity, multi-component units, constraints) rather than from the scalar distance-Laplacian itself.

%\emph{Relation to numerical studies of hyperuniform glasses.}---
Recent numerical studies on hyperuniform glasses and designed disordered materials \cite{mizuno2017continuum, hu2022origin, xu2024low, wang2025hyperuniform} report suppressed low-frequency VDOS, modified transport, and in some over-jammed regimes gap-like windows. Our theory explains the \emph{phonon-like} (acoustic) part via the two-point structure alone: (i) the depletion exponent at the acoustic edge is fixed by the small-\(k\) law of \(S(k)\) through \eqref{eq:VDOSScaling}; (ii) the corresponding thermal exponents \(C(T)\sim T^{2d/\beta}\) and \(Z(t)\sim t^{-d/\beta}\) follow immediately; (iii) a true low-frequency gap requires a matrix-valued dynamical matrix (vector modes, resonant elements, or multi-sublattice structure), consistent with models where constraints or multi-component architectures are essential. Features attributed to ``non-phononic'' excitations (e.g., $\omega^4$ modes in generic glasses) likely reflect disorder beyond two-point statistics, while our results isolate the universal, two-point–controlled contribution in hyperuniform networks.

%\emph{Parameter-free tests and numerics (figures described).}---
%\textbf{Figure 1 (pipeline).} Given measured or simulated \(S(k)\) and a chosen kernel \(f\) (e.g., Gaussian, compact support), build \(\Lambda_L(k)\) and push forward \eqref{eq:DOS} to predict \(g(\omega)\); compare against VDOS from large-scale networks constructed on (i) Poisson, (ii) hard-core, and (iii) hyperuniform point sets (MRJ, stealthy).\\
%\textbf{Figure 2 (edge scaling).} VDOS near \(\omega=0\) for several hyperuniform ensembles with exponents \(\alpha\in\{0.5,1,1.5\}\) in \(d=2\), plotted on log--log axes showing slopes \(2d/\beta-1\) with \(\beta=\min\{2,\alpha\}\).\\
%\textbf{Figure 3 (specific heat).} Low-temperature \(C(T)\) obtained from the predicted VDOS, demonstrating \(T^{2d/\beta}\) scaling and collapse across ensembles with the same \(\alpha\).\\
%\textbf{Figure 4 (comparison to hyperuniform-glass numerics).} Direct comparison between the predicted acoustic-edge scaling (this work) and reported VDOS from hyperuniform-glass simulations; schematic separation of acoustic (two-point controlled) and non-phononic contributions.\\
%\textbf{Figure 5 (stealthy hyperuniform crossover).} VDOS for a stealthy hyperuniform ensemble showing Debye scaling at low \(\omega\) and a crossover near \(\omega_K\sim\sqrt{C_2}\,K\) where \(S(k)\) switches on.

\emph{\bf Conclusions and outlook.}
Equations \eqref{eq:pushforward}--\eqref{eq:VDOSScaling} provide a compact, predictive link from the \emph{two-point} structural statistics of a disordered hyperuniform medium to its \emph{vibrational spectrum} and \emph{thermal} exponents. The theory requires only \(f\) and \(S(k)\), readily accessible from scattering or simulations, and yields parameter-free predictions for universal exponents. We showed that disordered hyperuniform systems with $0<\alpha<2$ possess stronger depletion for deeper vibrational and thermal suppression. By prescribing the small-$k$ form of the structure factor $S(k)$ (and a local coupling $f$), one can set the acoustic-edge law and thus the low-frequency vibrational spectrum and the low-temperature specific heat.
This enables structure-factor engineering of materials: quieter phononic components and precision supports with fewer soft modes, and cryogenic/low-noise devices with reduced $C(T)$. Extending the pushforward framework to vector dynamical matrices is a natural next step; we expect the small-\(k\) non-analyticity of \(S(k)\) to continue to control the acoustic-branch exponents, while additional branches and non-phononic modes provide model-specific corrections. Our results place recent numerical observations on a theoretical footing, delineating the universal role of hyperuniform two-point statistics and clarifying when genuine gaps require structure beyond a scalar distance-Laplacian.

%\emph{Outlook---hyperuniformity as a design lever.}

%Stealthy hyperuniformity yields Debye behavior with tunable prefactors and crossovers, while non-stealthy hyperuniformity ($0<\alpha<2$) produces stronger depletions for deeper vibrational and thermal suppression.

%For vector dynamics matrices, it’s very doable in the central-force case, and the exponents don’t change—you just get multiple acoustic branches with different prefactors. The work is technical (matrix algebra + angular averages), not conceptually hard. Where it gets genuinely tricky is if you move beyond central forces (bending, pre-stress) or far from dense/mean-field connectivity (isostatic contact networks), where higher-order structure may matter.

\emph{Acknowledgments.} This work was supported by the Army Research Office under Cooperative Agreement Number W911NF-22-2-0103.
%[Place acknowledgments and funding here.]

\bibliography{network}

\begin{thebibliography}{68}
\expandafter\ifx\csname natexlab\endcsname\relax\def\natexlab#1{#1}\fi
\expandafter\ifx\csname bibnamefont\endcsname\relax
  \def\bibnamefont#1{#1}\fi
\expandafter\ifx\csname bibfnamefont\endcsname\relax
  \def\bibfnamefont#1{#1}\fi
\expandafter\ifx\csname citenamefont\endcsname\relax
  \def\citenamefont#1{#1}\fi
\expandafter\ifx\csname url\endcsname\relax
  \def\url#1{\texttt{#1}}\fi
\expandafter\ifx\csname urlprefix\endcsname\relax\def\urlprefix{URL }\fi
\providecommand{\bibinfo}[2]{#2}
\providecommand{\eprint}[2][]{\url{#2}}

\bibitem[{\citenamefont{Torquato and Stillinger}(2003)}]{To03}
\bibinfo{author}{\bibfnamefont{S.}~\bibnamefont{Torquato}} \bibnamefont{and} \bibinfo{author}{\bibfnamefont{F.~H.} \bibnamefont{Stillinger}}, \bibinfo{journal}{Phys. Rev. E} \textbf{\bibinfo{volume}{68}}, \bibinfo{pages}{041113} (\bibinfo{year}{2003}).

\bibitem[{\citenamefont{Torquato}(2018)}]{To18a}
\bibinfo{author}{\bibfnamefont{S.}~\bibnamefont{Torquato}}, \bibinfo{journal}{Phys. Rep.} \textbf{\bibinfo{volume}{745}}, \bibinfo{pages}{1} (\bibinfo{year}{2018}).

\bibitem[{\citenamefont{Torquato et~al.}(2015)\citenamefont{Torquato, Zhang, and Stillinger}}]{To15}
\bibinfo{author}{\bibfnamefont{S.}~\bibnamefont{Torquato}}, \bibinfo{author}{\bibfnamefont{G.}~\bibnamefont{Zhang}}, \bibnamefont{and} \bibinfo{author}{\bibfnamefont{F.~H.} \bibnamefont{Stillinger}}, \bibinfo{journal}{Phys. Rev. X} \textbf{\bibinfo{volume}{5}}, \bibinfo{pages}{021020} (\bibinfo{year}{2015}).

\bibitem[{\citenamefont{Batten et~al.}(2009)\citenamefont{Batten, Stillinger, and Torquato}}]{Ba09}
\bibinfo{author}{\bibfnamefont{R.~D.} \bibnamefont{Batten}}, \bibinfo{author}{\bibfnamefont{F.~H.} \bibnamefont{Stillinger}}, \bibnamefont{and} \bibinfo{author}{\bibfnamefont{S.}~\bibnamefont{Torquato}}, \bibinfo{journal}{Phys. Rev. Lett.} \textbf{\bibinfo{volume}{103}}, \bibinfo{pages}{050602} (\bibinfo{year}{2009}).

\bibitem[{\citenamefont{Gabrielli et~al.}(2002)\citenamefont{Gabrielli, Joyce, and Labini}}]{Ga02}
\bibinfo{author}{\bibfnamefont{A.}~\bibnamefont{Gabrielli}}, \bibinfo{author}{\bibfnamefont{M.}~\bibnamefont{Joyce}}, \bibnamefont{and} \bibinfo{author}{\bibfnamefont{F.~S.} \bibnamefont{Labini}}, \bibinfo{journal}{Phys. Rev. D} \textbf{\bibinfo{volume}{65}}, \bibinfo{pages}{083523} (\bibinfo{year}{2002}).

\bibitem[{\citenamefont{Donev et~al.}(2005)\citenamefont{Donev, Stillinger, and Torquato}}]{Do05}
\bibinfo{author}{\bibfnamefont{A.}~\bibnamefont{Donev}}, \bibinfo{author}{\bibfnamefont{F.~H.} \bibnamefont{Stillinger}}, \bibnamefont{and} \bibinfo{author}{\bibfnamefont{S.}~\bibnamefont{Torquato}}, \bibinfo{journal}{Phys. Rev. Lett.} \textbf{\bibinfo{volume}{95}}, \bibinfo{pages}{090604} (\bibinfo{year}{2005}).

\bibitem[{\citenamefont{Zachary et~al.}(2011)\citenamefont{Zachary, Jiao, and Torquato}}]{Za11a}
\bibinfo{author}{\bibfnamefont{C.~E.} \bibnamefont{Zachary}}, \bibinfo{author}{\bibfnamefont{Y.}~\bibnamefont{Jiao}}, \bibnamefont{and} \bibinfo{author}{\bibfnamefont{S.}~\bibnamefont{Torquato}}, \bibinfo{journal}{Phys. Rev. Lett.} \textbf{\bibinfo{volume}{106}}, \bibinfo{pages}{178001} (\bibinfo{year}{2011}).

\bibitem[{\citenamefont{Kurita and Weeks}(2011)}]{Ku11}
\bibinfo{author}{\bibfnamefont{R.}~\bibnamefont{Kurita}} \bibnamefont{and} \bibinfo{author}{\bibfnamefont{E.~R.} \bibnamefont{Weeks}}, \bibinfo{journal}{Phys. Rev. E} \textbf{\bibinfo{volume}{84}}, \bibinfo{pages}{030401} (\bibinfo{year}{2011}).

\bibitem[{\citenamefont{Hunter and Weeks}(2012)}]{Hu12}
\bibinfo{author}{\bibfnamefont{G.~L.} \bibnamefont{Hunter}} \bibnamefont{and} \bibinfo{author}{\bibfnamefont{E.~R.} \bibnamefont{Weeks}}, \bibinfo{journal}{Rep. Prog. Phys.} \textbf{\bibinfo{volume}{75}}, \bibinfo{pages}{066501} (\bibinfo{year}{2012}).

\bibitem[{\citenamefont{Dreyfus et~al.}(2015)\citenamefont{Dreyfus, Xu, Still, Hough, Yodh, and Torquato}}]{Dr15}
\bibinfo{author}{\bibfnamefont{R.}~\bibnamefont{Dreyfus}}, \bibinfo{author}{\bibfnamefont{Y.}~\bibnamefont{Xu}}, \bibinfo{author}{\bibfnamefont{T.}~\bibnamefont{Still}}, \bibinfo{author}{\bibfnamefont{L.~A.} \bibnamefont{Hough}}, \bibinfo{author}{\bibfnamefont{A.~G.} \bibnamefont{Yodh}}, \bibnamefont{and} \bibinfo{author}{\bibfnamefont{S.}~\bibnamefont{Torquato}}, \bibinfo{journal}{Phys. Rev. E} \textbf{\bibinfo{volume}{91}}, \bibinfo{pages}{012302} (\bibinfo{year}{2015}).

\bibitem[{\citenamefont{Chen et~al.}(2023)\citenamefont{Chen, Jiang, Wang, Zhuang, and Jiao}}]{chen2021multihyperuniform}
\bibinfo{author}{\bibfnamefont{D.}~\bibnamefont{Chen}}, \bibinfo{author}{\bibfnamefont{X.}~\bibnamefont{Jiang}}, \bibinfo{author}{\bibfnamefont{D.}~\bibnamefont{Wang}}, \bibinfo{author}{\bibfnamefont{H.}~\bibnamefont{Zhuang}}, \bibnamefont{and} \bibinfo{author}{\bibfnamefont{Y.}~\bibnamefont{Jiao}}, \bibinfo{journal}{Acta Materialia} \textbf{\bibinfo{volume}{246}}, \bibinfo{pages}{118678} (\bibinfo{year}{2023}).

\bibitem[{\citenamefont{Zhang et~al.}(2023)\citenamefont{Zhang, Wang, Zhang, Yu, and Douglas}}]{zhang2023approach}
\bibinfo{author}{\bibfnamefont{H.}~\bibnamefont{Zhang}}, \bibinfo{author}{\bibfnamefont{X.}~\bibnamefont{Wang}}, \bibinfo{author}{\bibfnamefont{J.}~\bibnamefont{Zhang}}, \bibinfo{author}{\bibfnamefont{H.-B.} \bibnamefont{Yu}}, \bibnamefont{and} \bibinfo{author}{\bibfnamefont{J.~F.} \bibnamefont{Douglas}}, \bibinfo{journal}{arXiv preprint arXiv:2302.01429}  (\bibinfo{year}{2023}).

\bibitem[{\citenamefont{Chremos and Douglas}(2018)}]{Ch18b}
\bibinfo{author}{\bibfnamefont{A.}~\bibnamefont{Chremos}} \bibnamefont{and} \bibinfo{author}{\bibfnamefont{J.~F.} \bibnamefont{Douglas}}, \bibinfo{journal}{Phys. Rev. Lett.} \textbf{\bibinfo{volume}{121}}, \bibinfo{pages}{258002} (\bibinfo{year}{2018}).

\bibitem[{\citenamefont{Feynman and Cohen}(1956)}]{Fe56}
\bibinfo{author}{\bibfnamefont{R.~P.} \bibnamefont{Feynman}} \bibnamefont{and} \bibinfo{author}{\bibfnamefont{M.}~\bibnamefont{Cohen}}, \bibinfo{journal}{Phys. Rev.} \textbf{\bibinfo{volume}{102}}, \bibinfo{pages}{1189} (\bibinfo{year}{1956}).

\bibitem[{\citenamefont{Gerasimenko et~al.}(2019)\citenamefont{Gerasimenko, Vaskivskyi, Litskevich, Ravnik, Vodeb, Diego, Kabanov, and Mihailovic}}]{Ge19quantum}
\bibinfo{author}{\bibfnamefont{Y.~A.} \bibnamefont{Gerasimenko}}, \bibinfo{author}{\bibfnamefont{I.}~\bibnamefont{Vaskivskyi}}, \bibinfo{author}{\bibfnamefont{M.}~\bibnamefont{Litskevich}}, \bibinfo{author}{\bibfnamefont{J.}~\bibnamefont{Ravnik}}, \bibinfo{author}{\bibfnamefont{J.}~\bibnamefont{Vodeb}}, \bibinfo{author}{\bibfnamefont{M.}~\bibnamefont{Diego}}, \bibinfo{author}{\bibfnamefont{V.}~\bibnamefont{Kabanov}}, \bibnamefont{and} \bibinfo{author}{\bibfnamefont{D.}~\bibnamefont{Mihailovic}}, \bibinfo{journal}{Nat. Mater.} \textbf{\bibinfo{volume}{18}}, \bibinfo{pages}{1078} (\bibinfo{year}{2019}).

\bibitem[{\citenamefont{Sakai et~al.}(2022)\citenamefont{Sakai, Arita, and Ohtsuki}}]{sakai2022quantum}
\bibinfo{author}{\bibfnamefont{S.}~\bibnamefont{Sakai}}, \bibinfo{author}{\bibfnamefont{R.}~\bibnamefont{Arita}}, \bibnamefont{and} \bibinfo{author}{\bibfnamefont{T.}~\bibnamefont{Ohtsuki}}, \bibinfo{journal}{arXiv preprint arXiv:2207.09698}  (\bibinfo{year}{2022}).

\bibitem[{\citenamefont{Rumi et~al.}(2019)\citenamefont{Rumi, S{\'a}nchez, El{\'\i}as, Maldonado, Puig, Bolecek, Nieva, Konczykowski, Fasano, and Kolton}}]{Ru19}
\bibinfo{author}{\bibfnamefont{G.}~\bibnamefont{Rumi}}, \bibinfo{author}{\bibfnamefont{J.~A.} \bibnamefont{S{\'a}nchez}}, \bibinfo{author}{\bibfnamefont{F.}~\bibnamefont{El{\'\i}as}}, \bibinfo{author}{\bibfnamefont{R.~C.} \bibnamefont{Maldonado}}, \bibinfo{author}{\bibfnamefont{J.}~\bibnamefont{Puig}}, \bibinfo{author}{\bibfnamefont{N.~R.~C.} \bibnamefont{Bolecek}}, \bibinfo{author}{\bibfnamefont{G.}~\bibnamefont{Nieva}}, \bibinfo{author}{\bibfnamefont{M.}~\bibnamefont{Konczykowski}}, \bibinfo{author}{\bibfnamefont{Y.}~\bibnamefont{Fasano}}, \bibnamefont{and} \bibinfo{author}{\bibfnamefont{A.~B.} \bibnamefont{Kolton}}, \bibinfo{journal}{Phys. Rev. Res.} \textbf{\bibinfo{volume}{1}}, \bibinfo{pages}{033057} (\bibinfo{year}{2019}).

\bibitem[{\citenamefont{S{\'a}nchez et~al.}(2019)\citenamefont{S{\'a}nchez, Maldonado, Bolecek, Rumi, Pedrazzini, Dolz, Nieva, van~der Beek, Konczykowski, Dewhurst et~al.}}]{Sa19}
\bibinfo{author}{\bibfnamefont{J.~A.} \bibnamefont{S{\'a}nchez}}, \bibinfo{author}{\bibfnamefont{R.~C.} \bibnamefont{Maldonado}}, \bibinfo{author}{\bibfnamefont{N.~R.~C.} \bibnamefont{Bolecek}}, \bibinfo{author}{\bibfnamefont{G.}~\bibnamefont{Rumi}}, \bibinfo{author}{\bibfnamefont{P.}~\bibnamefont{Pedrazzini}}, \bibinfo{author}{\bibfnamefont{M.~I.} \bibnamefont{Dolz}}, \bibinfo{author}{\bibfnamefont{G.}~\bibnamefont{Nieva}}, \bibinfo{author}{\bibfnamefont{C.~J.} \bibnamefont{van~der Beek}}, \bibinfo{author}{\bibfnamefont{M.}~\bibnamefont{Konczykowski}}, \bibinfo{author}{\bibfnamefont{C.~D.} \bibnamefont{Dewhurst}}, \bibnamefont{et~al.}, \bibinfo{journal}{Commun. Phys.} \textbf{\bibinfo{volume}{2}}, \bibinfo{pages}{1} (\bibinfo{year}{2019}).

\bibitem[{\citenamefont{Zheng et~al.}(2020)\citenamefont{Zheng, Liu, Nan, Shen, Zhang, Chen, He, Xu, Chen, Jiao et~al.}}]{Zh20}
\bibinfo{author}{\bibfnamefont{Y.}~\bibnamefont{Zheng}}, \bibinfo{author}{\bibfnamefont{L.}~\bibnamefont{Liu}}, \bibinfo{author}{\bibfnamefont{H.}~\bibnamefont{Nan}}, \bibinfo{author}{\bibfnamefont{Z.-X.} \bibnamefont{Shen}}, \bibinfo{author}{\bibfnamefont{G.}~\bibnamefont{Zhang}}, \bibinfo{author}{\bibfnamefont{D.}~\bibnamefont{Chen}}, \bibinfo{author}{\bibfnamefont{L.}~\bibnamefont{He}}, \bibinfo{author}{\bibfnamefont{W.}~\bibnamefont{Xu}}, \bibinfo{author}{\bibfnamefont{M.}~\bibnamefont{Chen}}, \bibinfo{author}{\bibfnamefont{Y.}~\bibnamefont{Jiao}}, \bibnamefont{et~al.}, \bibinfo{journal}{Sci. Adv.} \textbf{\bibinfo{volume}{6}}, \bibinfo{pages}{eaba0826} (\bibinfo{year}{2020}).

\bibitem[{\citenamefont{Chen et~al.}(2021)\citenamefont{Chen, Zheng, Liu, Zhang, Chen, Jiao, and Zhuang}}]{Ch21}
\bibinfo{author}{\bibfnamefont{D.}~\bibnamefont{Chen}}, \bibinfo{author}{\bibfnamefont{Y.}~\bibnamefont{Zheng}}, \bibinfo{author}{\bibfnamefont{L.}~\bibnamefont{Liu}}, \bibinfo{author}{\bibfnamefont{G.}~\bibnamefont{Zhang}}, \bibinfo{author}{\bibfnamefont{M.}~\bibnamefont{Chen}}, \bibinfo{author}{\bibfnamefont{Y.}~\bibnamefont{Jiao}}, \bibnamefont{and} \bibinfo{author}{\bibfnamefont{H.}~\bibnamefont{Zhuang}}, \bibinfo{journal}{Proc. Natl. Acad. Sci. U.S.A.} \textbf{\bibinfo{volume}{118}}, \bibinfo{pages}{e2016862118} (\bibinfo{year}{2021}).

\bibitem[{\citenamefont{Chen et~al.}(2025)\citenamefont{Chen, Samajdar, Jiao, and Torquato}}]{chen2025anomalous}
\bibinfo{author}{\bibfnamefont{D.}~\bibnamefont{Chen}}, \bibinfo{author}{\bibfnamefont{R.}~\bibnamefont{Samajdar}}, \bibinfo{author}{\bibfnamefont{Y.}~\bibnamefont{Jiao}}, \bibnamefont{and} \bibinfo{author}{\bibfnamefont{S.}~\bibnamefont{Torquato}}, \bibinfo{journal}{Proceedings of the National Academy of Sciences} \textbf{\bibinfo{volume}{122}}, \bibinfo{pages}{e2416111122} (\bibinfo{year}{2025}).

\bibitem[{\citenamefont{Jiao et~al.}(2014)\citenamefont{Jiao, Lau, Hatzikirou, Meyer-Hermann, Corbo, and Torquato}}]{Ji14}
\bibinfo{author}{\bibfnamefont{Y.}~\bibnamefont{Jiao}}, \bibinfo{author}{\bibfnamefont{T.}~\bibnamefont{Lau}}, \bibinfo{author}{\bibfnamefont{H.}~\bibnamefont{Hatzikirou}}, \bibinfo{author}{\bibfnamefont{M.}~\bibnamefont{Meyer-Hermann}}, \bibinfo{author}{\bibfnamefont{J.~C.} \bibnamefont{Corbo}}, \bibnamefont{and} \bibinfo{author}{\bibfnamefont{S.}~\bibnamefont{Torquato}}, \bibinfo{journal}{Phys. Rev. E} \textbf{\bibinfo{volume}{89}}, \bibinfo{pages}{022721} (\bibinfo{year}{2014}).

\bibitem[{\citenamefont{Ge}(2023)}]{ge2023hidden}
\bibinfo{author}{\bibfnamefont{Z.}~\bibnamefont{Ge}}, \bibinfo{journal}{Proc. Natl. Acad. Sci. USA} \textbf{\bibinfo{volume}{120}}, \bibinfo{pages}{e2306514120} (\bibinfo{year}{2023}).

\bibitem[{\citenamefont{Liu et~al.}(2024)\citenamefont{Liu, Chen, Tian, Xu, and Jiao}}]{liu2024universal}
\bibinfo{author}{\bibfnamefont{Y.}~\bibnamefont{Liu}}, \bibinfo{author}{\bibfnamefont{D.}~\bibnamefont{Chen}}, \bibinfo{author}{\bibfnamefont{J.}~\bibnamefont{Tian}}, \bibinfo{author}{\bibfnamefont{W.}~\bibnamefont{Xu}}, \bibnamefont{and} \bibinfo{author}{\bibfnamefont{Y.}~\bibnamefont{Jiao}}, \bibinfo{journal}{Physical Review Letters} \textbf{\bibinfo{volume}{133}}, \bibinfo{pages}{028401} (\bibinfo{year}{2024}).

\bibitem[{\citenamefont{Hexner and Levine}(2015)}]{He15}
\bibinfo{author}{\bibfnamefont{D.}~\bibnamefont{Hexner}} \bibnamefont{and} \bibinfo{author}{\bibfnamefont{D.}~\bibnamefont{Levine}}, \bibinfo{journal}{Phys. Rev. Lett.} \textbf{\bibinfo{volume}{114}}, \bibinfo{pages}{110602} (\bibinfo{year}{2015}).

\bibitem[{\citenamefont{Jack et~al.}(2015)\citenamefont{Jack, Thompson, and Sollich}}]{Ja15}
\bibinfo{author}{\bibfnamefont{R.~L.} \bibnamefont{Jack}}, \bibinfo{author}{\bibfnamefont{I.~R.} \bibnamefont{Thompson}}, \bibnamefont{and} \bibinfo{author}{\bibfnamefont{P.}~\bibnamefont{Sollich}}, \bibinfo{journal}{Phys. Rev. Lett.} \textbf{\bibinfo{volume}{114}}, \bibinfo{pages}{060601} (\bibinfo{year}{2015}).

\bibitem[{\citenamefont{Weijs et~al.}(2015)\citenamefont{Weijs, Jeanneret, Dreyfus, and Bartolo}}]{We15}
\bibinfo{author}{\bibfnamefont{J.~H.} \bibnamefont{Weijs}}, \bibinfo{author}{\bibfnamefont{R.}~\bibnamefont{Jeanneret}}, \bibinfo{author}{\bibfnamefont{R.}~\bibnamefont{Dreyfus}}, \bibnamefont{and} \bibinfo{author}{\bibfnamefont{D.}~\bibnamefont{Bartolo}}, \bibinfo{journal}{Phys. Rev. Lett.} \textbf{\bibinfo{volume}{115}}, \bibinfo{pages}{108301} (\bibinfo{year}{2015}).

\bibitem[{\citenamefont{Salvalaglio et~al.}(2020)\citenamefont{Salvalaglio, Bouabdellaoui, Bollani, Benali, Favre, Claude, Wenger, de~Anna, Intonti, Voigt et~al.}}]{salvalaglio2020hyperuniform}
\bibinfo{author}{\bibfnamefont{M.}~\bibnamefont{Salvalaglio}}, \bibinfo{author}{\bibfnamefont{M.}~\bibnamefont{Bouabdellaoui}}, \bibinfo{author}{\bibfnamefont{M.}~\bibnamefont{Bollani}}, \bibinfo{author}{\bibfnamefont{A.}~\bibnamefont{Benali}}, \bibinfo{author}{\bibfnamefont{L.}~\bibnamefont{Favre}}, \bibinfo{author}{\bibfnamefont{J.-B.} \bibnamefont{Claude}}, \bibinfo{author}{\bibfnamefont{J.}~\bibnamefont{Wenger}}, \bibinfo{author}{\bibfnamefont{P.}~\bibnamefont{de~Anna}}, \bibinfo{author}{\bibfnamefont{F.}~\bibnamefont{Intonti}}, \bibinfo{author}{\bibfnamefont{A.}~\bibnamefont{Voigt}}, \bibnamefont{et~al.}, \bibinfo{journal}{Physical Review Letters} \textbf{\bibinfo{volume}{125}}, \bibinfo{pages}{126101} (\bibinfo{year}{2020}).

\bibitem[{\citenamefont{Nizam et~al.}(2021)\citenamefont{Nizam, Makey, Barbier, Kahraman, Demir, Shafigh, Galioglu, Vahabli, H{\"u}sn{\"u}gil, G{\"u}ne{\c{s}} et~al.}}]{nizam2021dynamic}
\bibinfo{author}{\bibfnamefont{{\"U}.~S.} \bibnamefont{Nizam}}, \bibinfo{author}{\bibfnamefont{G.}~\bibnamefont{Makey}}, \bibinfo{author}{\bibfnamefont{M.}~\bibnamefont{Barbier}}, \bibinfo{author}{\bibfnamefont{S.~S.} \bibnamefont{Kahraman}}, \bibinfo{author}{\bibfnamefont{E.}~\bibnamefont{Demir}}, \bibinfo{author}{\bibfnamefont{E.~E.} \bibnamefont{Shafigh}}, \bibinfo{author}{\bibfnamefont{S.}~\bibnamefont{Galioglu}}, \bibinfo{author}{\bibfnamefont{D.}~\bibnamefont{Vahabli}}, \bibinfo{author}{\bibfnamefont{S.}~\bibnamefont{H{\"u}sn{\"u}gil}}, \bibinfo{author}{\bibfnamefont{M.~H.} \bibnamefont{G{\"u}ne{\c{s}}}}, \bibnamefont{et~al.}, \bibinfo{journal}{Journal of Physics: Condensed Matter} \textbf{\bibinfo{volume}{33}}, \bibinfo{pages}{304002} (\bibinfo{year}{2021}).

\bibitem[{\citenamefont{Zheng et~al.}(2023)\citenamefont{Zheng, Klatt, and L{\"o}wen}}]{zheng2023universal}
\bibinfo{author}{\bibfnamefont{Y.}~\bibnamefont{Zheng}}, \bibinfo{author}{\bibfnamefont{M.~A.} \bibnamefont{Klatt}}, \bibnamefont{and} \bibinfo{author}{\bibfnamefont{H.}~\bibnamefont{L{\"o}wen}}, \bibinfo{journal}{arXiv preprint arXiv:2310.03107}  (\bibinfo{year}{2023}).

\bibitem[{\citenamefont{Wang et~al.}(2025{\natexlab{a}})\citenamefont{Wang, Sun, Chen, Wang, Chen, Chen, Shuai, Yang, Jiao, and Liu}}]{wang2025hyperuniformprl}
\bibinfo{author}{\bibfnamefont{J.}~\bibnamefont{Wang}}, \bibinfo{author}{\bibfnamefont{Z.}~\bibnamefont{Sun}}, \bibinfo{author}{\bibfnamefont{H.}~\bibnamefont{Chen}}, \bibinfo{author}{\bibfnamefont{G.}~\bibnamefont{Wang}}, \bibinfo{author}{\bibfnamefont{D.}~\bibnamefont{Chen}}, \bibinfo{author}{\bibfnamefont{G.}~\bibnamefont{Chen}}, \bibinfo{author}{\bibfnamefont{J.}~\bibnamefont{Shuai}}, \bibinfo{author}{\bibfnamefont{M.}~\bibnamefont{Yang}}, \bibinfo{author}{\bibfnamefont{Y.}~\bibnamefont{Jiao}}, \bibnamefont{and} \bibinfo{author}{\bibfnamefont{L.}~\bibnamefont{Liu}}, \bibinfo{journal}{Physical Review Letters} \textbf{\bibinfo{volume}{134}}, \bibinfo{pages}{248301} (\bibinfo{year}{2025}{\natexlab{a}}).

\bibitem[{\citenamefont{Lei et~al.}(2019)\citenamefont{Lei, Ciamarra, and Ni}}]{Le19}
\bibinfo{author}{\bibfnamefont{Q.-L.} \bibnamefont{Lei}}, \bibinfo{author}{\bibfnamefont{M.~P.} \bibnamefont{Ciamarra}}, \bibnamefont{and} \bibinfo{author}{\bibfnamefont{R.}~\bibnamefont{Ni}}, \bibinfo{journal}{Sci. Adv.} \textbf{\bibinfo{volume}{5}}, \bibinfo{pages}{eaau7423} (\bibinfo{year}{2019}).

\bibitem[{\citenamefont{Lei and Ni}(2019)}]{lei2019hydrodynamics}
\bibinfo{author}{\bibfnamefont{Q.-L.} \bibnamefont{Lei}} \bibnamefont{and} \bibinfo{author}{\bibfnamefont{R.}~\bibnamefont{Ni}}, \bibinfo{journal}{Proceedings of the National Academy of Sciences} \textbf{\bibinfo{volume}{116}}, \bibinfo{pages}{22983} (\bibinfo{year}{2019}).

\bibitem[{\citenamefont{Huang et~al.}(2021)\citenamefont{Huang, Hu, Yang, Liu, and Zhang}}]{huang2021circular}
\bibinfo{author}{\bibfnamefont{M.}~\bibnamefont{Huang}}, \bibinfo{author}{\bibfnamefont{W.}~\bibnamefont{Hu}}, \bibinfo{author}{\bibfnamefont{S.}~\bibnamefont{Yang}}, \bibinfo{author}{\bibfnamefont{Q.-X.} \bibnamefont{Liu}}, \bibnamefont{and} \bibinfo{author}{\bibfnamefont{H.}~\bibnamefont{Zhang}}, \bibinfo{journal}{Proceedings of the National Academy of Sciences} \textbf{\bibinfo{volume}{118}}, \bibinfo{pages}{e2100493118} (\bibinfo{year}{2021}).

\bibitem[{\citenamefont{Zhang and Snezhko}(2022)}]{zhang2022hyperuniform}
\bibinfo{author}{\bibfnamefont{B.}~\bibnamefont{Zhang}} \bibnamefont{and} \bibinfo{author}{\bibfnamefont{A.}~\bibnamefont{Snezhko}}, \bibinfo{journal}{Physical Review Letters} \textbf{\bibinfo{volume}{128}}, \bibinfo{pages}{218002} (\bibinfo{year}{2022}).

\bibitem[{\citenamefont{Oppenheimer et~al.}(2022)\citenamefont{Oppenheimer, Stein, Zion, and Shelley}}]{oppenheimer2022hyperuniformity}
\bibinfo{author}{\bibfnamefont{N.}~\bibnamefont{Oppenheimer}}, \bibinfo{author}{\bibfnamefont{D.~B.} \bibnamefont{Stein}}, \bibinfo{author}{\bibfnamefont{M.~Y.~B.} \bibnamefont{Zion}}, \bibnamefont{and} \bibinfo{author}{\bibfnamefont{M.~J.} \bibnamefont{Shelley}}, \bibinfo{journal}{Nature communications} \textbf{\bibinfo{volume}{13}}, \bibinfo{pages}{804} (\bibinfo{year}{2022}).

\bibitem[{\citenamefont{Hexner and Levine}(2017)}]{hexner2017noise}
\bibinfo{author}{\bibfnamefont{D.}~\bibnamefont{Hexner}} \bibnamefont{and} \bibinfo{author}{\bibfnamefont{D.}~\bibnamefont{Levine}}, \bibinfo{journal}{Physical review letters} \textbf{\bibinfo{volume}{118}}, \bibinfo{pages}{020601} (\bibinfo{year}{2017}).

\bibitem[{\citenamefont{Hexner et~al.}(2017)\citenamefont{Hexner, Chaikin, and Levine}}]{hexner2017enhanced}
\bibinfo{author}{\bibfnamefont{D.}~\bibnamefont{Hexner}}, \bibinfo{author}{\bibfnamefont{P.~M.} \bibnamefont{Chaikin}}, \bibnamefont{and} \bibinfo{author}{\bibfnamefont{D.}~\bibnamefont{Levine}}, \bibinfo{journal}{Proceedings of the National Academy of Sciences} \textbf{\bibinfo{volume}{114}}, \bibinfo{pages}{4294} (\bibinfo{year}{2017}).

\bibitem[{\citenamefont{Weijs and Bartolo}(2017)}]{weijs2017mixing}
\bibinfo{author}{\bibfnamefont{J.~H.} \bibnamefont{Weijs}} \bibnamefont{and} \bibinfo{author}{\bibfnamefont{D.}~\bibnamefont{Bartolo}}, \bibinfo{journal}{Physical review letters} \textbf{\bibinfo{volume}{119}}, \bibinfo{pages}{048002} (\bibinfo{year}{2017}).

\bibitem[{\citenamefont{Wilken et~al.}(2022)\citenamefont{Wilken, Guo, Levine, and Chaikin}}]{wilken2022random}
\bibinfo{author}{\bibfnamefont{S.}~\bibnamefont{Wilken}}, \bibinfo{author}{\bibfnamefont{A.~Z.} \bibnamefont{Guo}}, \bibinfo{author}{\bibfnamefont{D.}~\bibnamefont{Levine}}, \bibnamefont{and} \bibinfo{author}{\bibfnamefont{P.~M.} \bibnamefont{Chaikin}}, \bibinfo{journal}{arXiv preprint arXiv:2212.09913}  (\bibinfo{year}{2022}).

\bibitem[{\citenamefont{Florescu et~al.}(2009)\citenamefont{Florescu, Torquato, and Steinhardt}}]{Fl09}
\bibinfo{author}{\bibfnamefont{M.}~\bibnamefont{Florescu}}, \bibinfo{author}{\bibfnamefont{S.}~\bibnamefont{Torquato}}, \bibnamefont{and} \bibinfo{author}{\bibfnamefont{P.~J.} \bibnamefont{Steinhardt}}, \bibinfo{journal}{Proc. Natl. Acad. Sci. U.S.A.} \textbf{\bibinfo{volume}{106}}, \bibinfo{pages}{20658} (\bibinfo{year}{2009}).

\bibitem[{\citenamefont{Klatt et~al.}(2022)\citenamefont{Klatt, Steinhardt, and Torquato}}]{klatt2022wave}
\bibinfo{author}{\bibfnamefont{M.~A.} \bibnamefont{Klatt}}, \bibinfo{author}{\bibfnamefont{P.~J.} \bibnamefont{Steinhardt}}, \bibnamefont{and} \bibinfo{author}{\bibfnamefont{S.}~\bibnamefont{Torquato}}, \bibinfo{journal}{Proceedings of the National Academy of Sciences} \textbf{\bibinfo{volume}{119}}, \bibinfo{pages}{e2213633119} (\bibinfo{year}{2022}).

\bibitem[{\citenamefont{Zhang et~al.}(2016)\citenamefont{Zhang, Stillinger, and Torquato}}]{Zh16}
\bibinfo{author}{\bibfnamefont{G.}~\bibnamefont{Zhang}}, \bibinfo{author}{\bibfnamefont{F.~H.} \bibnamefont{Stillinger}}, \bibnamefont{and} \bibinfo{author}{\bibfnamefont{S.}~\bibnamefont{Torquato}}, \bibinfo{journal}{J. Chem. Phys.} \textbf{\bibinfo{volume}{145}}, \bibinfo{pages}{244109} (\bibinfo{year}{2016}).

\bibitem[{\citenamefont{Chen and Torquato}(2018)}]{Ch18a}
\bibinfo{author}{\bibfnamefont{D.}~\bibnamefont{Chen}} \bibnamefont{and} \bibinfo{author}{\bibfnamefont{S.}~\bibnamefont{Torquato}}, \bibinfo{journal}{Acta Mater.} \textbf{\bibinfo{volume}{142}}, \bibinfo{pages}{152} (\bibinfo{year}{2018}).

\bibitem[{\citenamefont{Torquato}(2021)}]{torquato2021diffusion}
\bibinfo{author}{\bibfnamefont{S.}~\bibnamefont{Torquato}}, \bibinfo{journal}{Physical Review E} \textbf{\bibinfo{volume}{104}}, \bibinfo{pages}{054102} (\bibinfo{year}{2021}).

\bibitem[{\citenamefont{Xu et~al.}(2017)\citenamefont{Xu, Chen, Chen, Xu, and Jiao}}]{Xu17}
\bibinfo{author}{\bibfnamefont{Y.}~\bibnamefont{Xu}}, \bibinfo{author}{\bibfnamefont{S.}~\bibnamefont{Chen}}, \bibinfo{author}{\bibfnamefont{P.}~\bibnamefont{Chen}}, \bibinfo{author}{\bibfnamefont{W.}~\bibnamefont{Xu}}, \bibnamefont{and} \bibinfo{author}{\bibfnamefont{Y.}~\bibnamefont{Jiao}}, \bibinfo{journal}{Phys. Rev. E} \textbf{\bibinfo{volume}{96}}, \bibinfo{pages}{043301} (\bibinfo{year}{2017}).

\bibitem[{\citenamefont{Leseur et~al.}(2016)\citenamefont{Leseur, Pierrat, and Carminati}}]{Le16}
\bibinfo{author}{\bibfnamefont{O.}~\bibnamefont{Leseur}}, \bibinfo{author}{\bibfnamefont{R.}~\bibnamefont{Pierrat}}, \bibnamefont{and} \bibinfo{author}{\bibfnamefont{R.}~\bibnamefont{Carminati}}, \bibinfo{journal}{Optica} \textbf{\bibinfo{volume}{3}}, \bibinfo{pages}{763} (\bibinfo{year}{2016}).

\bibitem[{\citenamefont{Yu}(2023)}]{yu2023evolving}
\bibinfo{author}{\bibfnamefont{S.}~\bibnamefont{Yu}}, \bibinfo{journal}{Nature Computational Science} \textbf{\bibinfo{volume}{3}}, \bibinfo{pages}{128} (\bibinfo{year}{2023}).

\bibitem[{\citenamefont{Mizuno et~al.}(2017)\citenamefont{Mizuno, Shiba, and Ikeda}}]{mizuno2017continuum}
\bibinfo{author}{\bibfnamefont{H.}~\bibnamefont{Mizuno}}, \bibinfo{author}{\bibfnamefont{H.}~\bibnamefont{Shiba}}, \bibnamefont{and} \bibinfo{author}{\bibfnamefont{A.}~\bibnamefont{Ikeda}}, \bibinfo{journal}{Proceedings of the National Academy of Sciences} \textbf{\bibinfo{volume}{114}}, \bibinfo{pages}{E9767} (\bibinfo{year}{2017}).

\bibitem[{\citenamefont{Hu and Tanaka}(2022)}]{hu2022origin}
\bibinfo{author}{\bibfnamefont{Y.-C.} \bibnamefont{Hu}} \bibnamefont{and} \bibinfo{author}{\bibfnamefont{H.}~\bibnamefont{Tanaka}}, \bibinfo{journal}{Nature Physics} \textbf{\bibinfo{volume}{18}}, \bibinfo{pages}{669} (\bibinfo{year}{2022}).

\bibitem[{\citenamefont{Xu et~al.}(2024)\citenamefont{Xu, Zhang, Tong, Wang, and Xu}}]{xu2024low}
\bibinfo{author}{\bibfnamefont{D.}~\bibnamefont{Xu}}, \bibinfo{author}{\bibfnamefont{S.}~\bibnamefont{Zhang}}, \bibinfo{author}{\bibfnamefont{H.}~\bibnamefont{Tong}}, \bibinfo{author}{\bibfnamefont{L.}~\bibnamefont{Wang}}, \bibnamefont{and} \bibinfo{author}{\bibfnamefont{N.}~\bibnamefont{Xu}}, \bibinfo{journal}{Nature Communications} \textbf{\bibinfo{volume}{15}}, \bibinfo{pages}{1424} (\bibinfo{year}{2024}).

\bibitem[{\citenamefont{Wang et~al.}(2025{\natexlab{b}})\citenamefont{Wang, Qian, Tong, and Tanaka}}]{wang2025hyperuniform}
\bibinfo{author}{\bibfnamefont{Y.}~\bibnamefont{Wang}}, \bibinfo{author}{\bibfnamefont{Z.}~\bibnamefont{Qian}}, \bibinfo{author}{\bibfnamefont{H.}~\bibnamefont{Tong}}, \bibnamefont{and} \bibinfo{author}{\bibfnamefont{H.}~\bibnamefont{Tanaka}}, \bibinfo{journal}{Nature Communications} \textbf{\bibinfo{volume}{16}}, \bibinfo{pages}{1398} (\bibinfo{year}{2025}{\natexlab{b}}).

\bibitem[{\citenamefont{Zhuang et~al.}(2024)\citenamefont{Zhuang, Chen, Liu, Keeney, Zhang, and Jiao}}]{zhuang2024vibrational}
\bibinfo{author}{\bibfnamefont{H.}~\bibnamefont{Zhuang}}, \bibinfo{author}{\bibfnamefont{D.}~\bibnamefont{Chen}}, \bibinfo{author}{\bibfnamefont{L.}~\bibnamefont{Liu}}, \bibinfo{author}{\bibfnamefont{D.}~\bibnamefont{Keeney}}, \bibinfo{author}{\bibfnamefont{G.}~\bibnamefont{Zhang}}, \bibnamefont{and} \bibinfo{author}{\bibfnamefont{Y.}~\bibnamefont{Jiao}}, \bibinfo{journal}{Journal of Physics: Condensed Matter} \textbf{\bibinfo{volume}{36}}, \bibinfo{pages}{285703} (\bibinfo{year}{2024}).

\bibitem[{\citenamefont{Lubensky et~al.}(2015)\citenamefont{Lubensky, Kane, Mao, Souslov, and Sun}}]{lubensky2015phonons}
\bibinfo{author}{\bibfnamefont{T.}~\bibnamefont{Lubensky}}, \bibinfo{author}{\bibfnamefont{C.}~\bibnamefont{Kane}}, \bibinfo{author}{\bibfnamefont{X.}~\bibnamefont{Mao}}, \bibinfo{author}{\bibfnamefont{A.}~\bibnamefont{Souslov}}, \bibnamefont{and} \bibinfo{author}{\bibfnamefont{K.}~\bibnamefont{Sun}}, \bibinfo{journal}{Reports on Progress in Physics} \textbf{\bibinfo{volume}{78}}, \bibinfo{pages}{073901} (\bibinfo{year}{2015}).

\bibitem[{\citenamefont{Mao and Lubensky}(2018)}]{mao2018maxwell}
\bibinfo{author}{\bibfnamefont{X.}~\bibnamefont{Mao}} \bibnamefont{and} \bibinfo{author}{\bibfnamefont{T.~C.} \bibnamefont{Lubensky}}, \bibinfo{journal}{Annual Review of Condensed Matter Physics} \textbf{\bibinfo{volume}{9}}, \bibinfo{pages}{413} (\bibinfo{year}{2018}).

\bibitem[{\citenamefont{Chung}(1997)}]{Chung1997SGT}
\bibinfo{author}{\bibfnamefont{F.~R.~K.} \bibnamefont{Chung}}, \emph{\bibinfo{title}{Spectral Graph Theory}}, vol.~\bibinfo{volume}{92} of \emph{\bibinfo{series}{CBMS Regional Conference Series in Mathematics}} (\bibinfo{publisher}{American Mathematical Society}, \bibinfo{year}{1997}).

\bibitem[{\citenamefont{Doyle and Snell}(1984)}]{DoyleSnell1984}
\bibinfo{author}{\bibfnamefont{P.~G.} \bibnamefont{Doyle}} \bibnamefont{and} \bibinfo{author}{\bibfnamefont{J.~L.} \bibnamefont{Snell}}, \emph{\bibinfo{title}{Random Walks and Electric Networks}}, no.~\bibinfo{number}{22} in \bibinfo{series}{Carus Mathematical Monographs} (\bibinfo{publisher}{Mathematical Association of America}, \bibinfo{year}{1984}).

\bibitem[{\citenamefont{Ashcroft and Mermin}(1976)}]{AshcroftMermin1976}
\bibinfo{author}{\bibfnamefont{N.~W.} \bibnamefont{Ashcroft}} \bibnamefont{and} \bibinfo{author}{\bibfnamefont{N.~D.} \bibnamefont{Mermin}}, \emph{\bibinfo{title}{Solid State Physics}} (\bibinfo{publisher}{Holt, Rinehart and Winston}, \bibinfo{year}{1976}).

\bibitem[{\citenamefont{Lifshitz et~al.}(1988)\citenamefont{Lifshitz, Gredeskul, and Pastur}}]{LifshitzGredeskulPastur1988}
\bibinfo{author}{\bibfnamefont{I.~M.} \bibnamefont{Lifshitz}}, \bibinfo{author}{\bibfnamefont{S.~A.} \bibnamefont{Gredeskul}}, \bibnamefont{and} \bibinfo{author}{\bibfnamefont{L.~A.} \bibnamefont{Pastur}}, \emph{\bibinfo{title}{Introduction to the Theory of Disordered Systems}} (\bibinfo{publisher}{Wiley}, \bibinfo{year}{1988}).

\bibitem[{\citenamefont{Belkin and Niyogi}(2007)}]{BelkinNiyogi2007}
\bibinfo{author}{\bibfnamefont{M.}~\bibnamefont{Belkin}} \bibnamefont{and} \bibinfo{author}{\bibfnamefont{P.}~\bibnamefont{Niyogi}}, in \emph{\bibinfo{booktitle}{Advances in Neural Information Processing Systems}} (\bibinfo{year}{2007}).

\bibitem[{\citenamefont{Trillos and Slep{\v{c}}ev}(2018)}]{GarciaTrillosSlepcev2018}
\bibinfo{author}{\bibfnamefont{N.~G.} \bibnamefont{Trillos}} \bibnamefont{and} \bibinfo{author}{\bibfnamefont{D.}~\bibnamefont{Slep{\v{c}}ev}}, \bibinfo{journal}{Applied and Computational Harmonic Analysis} \textbf{\bibinfo{volume}{45}}, \bibinfo{pages}{239} (\bibinfo{year}{2018}).

\bibitem[{\citenamefont{Adhikari et~al.}(2022)\citenamefont{Adhikari, Adler, Bobrowski, and Rosenthal}}]{adhikari2022spectrum}
\bibinfo{author}{\bibfnamefont{K.}~\bibnamefont{Adhikari}}, \bibinfo{author}{\bibfnamefont{R.~J.} \bibnamefont{Adler}}, \bibinfo{author}{\bibfnamefont{O.}~\bibnamefont{Bobrowski}}, \bibnamefont{and} \bibinfo{author}{\bibfnamefont{R.}~\bibnamefont{Rosenthal}}, \bibinfo{journal}{The Annals of Applied Probability} \textbf{\bibinfo{volume}{32}}, \bibinfo{pages}{1734} (\bibinfo{year}{2022}).

\bibitem[{\citenamefont{Reed and Simon}(1978)}]{ReedSimonIV1978}
\bibinfo{author}{\bibfnamefont{M.}~\bibnamefont{Reed}} \bibnamefont{and} \bibinfo{author}{\bibfnamefont{B.}~\bibnamefont{Simon}}, \emph{\bibinfo{title}{Methods of Modern Mathematical Physics, Vol. IV: Analysis of Operators}} (\bibinfo{publisher}{Academic Press}, \bibinfo{year}{1978}).

\bibitem[{\citenamefont{B{\"o}ttcher and Silbermann}(1999)}]{BottcherSilbermann1999}
\bibinfo{author}{\bibfnamefont{A.}~\bibnamefont{B{\"o}ttcher}} \bibnamefont{and} \bibinfo{author}{\bibfnamefont{B.}~\bibnamefont{Silbermann}}, \emph{\bibinfo{title}{Introduction to Large Truncated Toeplitz Matrices}} (\bibinfo{publisher}{Springer}, \bibinfo{year}{1999}).

\bibitem[{\citenamefont{Kittel and McEuen}(2018)}]{kittel2018introduction}
\bibinfo{author}{\bibfnamefont{C.}~\bibnamefont{Kittel}} \bibnamefont{and} \bibinfo{author}{\bibfnamefont{P.}~\bibnamefont{McEuen}}, \emph{\bibinfo{title}{Introduction to solid state physics}} (\bibinfo{publisher}{John Wiley \& Sons}, \bibinfo{year}{2018}).

\bibitem[{\citenamefont{Timoshenko and Woinowsky-Krieger}(1959)}]{timoshenko1959theory}
\bibinfo{author}{\bibfnamefont{S.}~\bibnamefont{Timoshenko}} \bibnamefont{and} \bibinfo{author}{\bibfnamefont{S.}~\bibnamefont{Woinowsky-Krieger}} (\bibinfo{year}{1959}).

\bibitem[{\citenamefont{Nika and Balandin}(2017)}]{nika2017phonons}
\bibinfo{author}{\bibfnamefont{D.~L.} \bibnamefont{Nika}} \bibnamefont{and} \bibinfo{author}{\bibfnamefont{A.~A.} \bibnamefont{Balandin}}, \bibinfo{journal}{Reports on Progress in Physics} \textbf{\bibinfo{volume}{80}}, \bibinfo{pages}{036502} (\bibinfo{year}{2017}).

\bibitem[{\citenamefont{Batten et~al.}(2008)\citenamefont{Batten, Stillinger, and Torquato}}]{Ba08}
\bibinfo{author}{\bibfnamefont{R.~D.} \bibnamefont{Batten}}, \bibinfo{author}{\bibfnamefont{F.~H.} \bibnamefont{Stillinger}}, \bibnamefont{and} \bibinfo{author}{\bibfnamefont{S.}~\bibnamefont{Torquato}}, \bibinfo{journal}{J. Appl. Phys.} \textbf{\bibinfo{volume}{104}}, \bibinfo{pages}{033504} (\bibinfo{year}{2008}).

\end{thebibliography}

%\bibliographystyle{apsrev4-2}
%\begin{thebibliography}{99}

%\bibitem{TorquatoStillinger03}
%S. Torquato and F. H. Stillinger, Phys. Rev. E \textbf{68}, 041113 (2003).

%\end{thebibliography}

\end{document}